# Exotic in dense and cold nuclear matter.

*K.Mikhailov,A.Stavinskiy,G.Sharkov,V.Stolin(ITEP)*

*Abstract*
The way to create and to investigate a dense cold matter droplets in the laboratory is proposed. The reality of this approach are argued. Estimated possible statistic is large enough for detail study of the properties of such a matter.Mechanism of kinematical cooling of the droplet is clarified.Different types of trigger(selection criteria) are proposed to search for different kind of exotic.

## 1. Introduction

All reliable known baryons consists of three quarks, while pentaquarks(consisting of four quarks and one antiquark) are not forbidden by fundamental principes and there is theoretical models with pentaquarks, for example[1]. The pentaquarks are being searched for a long time without definite results. It looks like the cross sections for pentaquarks production is extrimely small if not zero. For example upper limit for pentaquark production cross section is 0.7nb according to CLAS experiments[2]. Similar sutuation, from experimental point of view, is for dibaryons. It is not a trivial task to find an effect with so small production cross section. Natural question in this context – is it possible to apply selection criteria(or special trigger) to reject significantly (for the orders of magnitude) ordinary baryons production without significant (ideally-any) rejection of the process under the study (pentaquark production). In another words, is it possible to create exotic conditions (rare process in the nature) in which pentaquark production would be natural and not supressed.

An example of that is being discussed in the astrophysics in context of so-called neutron(compact) stars. Due to the large baryonic density of neutron stars core neutron chemical potential $\mu_n$ became larger then hyperon production threshold $\mu_\Lambda$. From the relation $\mu_n=\mu_\Lambda$ one can conclude that such transformation would be at neutron Fermi-momentum $k_{Fn} \sim 3$ fm$^{-1}$; it corresponds to nuclear matter density two times larger than standard nuclear matter density $\rho_0 = 0.16$ fm$^{-3}$ [3].

We will discuss below in this article the possibility to create in the laboratory such conditions in which pentaquarks and other exotics could became significant part of baryons content in the produced system.

## 2. Basic conception.

Suppose we can create dense and cold quark matter (like one in the neutron star core or even denser) in the laboratory. We will consider the matter as sufficiently dense and cold if all possible due to Pauli exclusion principle quantun states are filled or close to be filled ($\delta r \delta p \sim h$). We will define $\delta p$ from the temperature of the matter as follows: $\delta p \sim \sqrt{2m_N T_0}$. In case if all quantun states for the light quarks are filled or close to be filled but there is no available energy for a transition into heavy quarks (s,c,b,t) the increase of the system density is possible due to its bosonization, for example in the form of diquarks. If diquarks becomes remarkable or may be even dominant constituent of the system, pentaquark production is not suppressed with respect to ordinary baryons from topological point of view. Moreover for purely diquark mediun ordinary baryons would become an exotic with respect to pentaquarks and dibaryons(fig.1).

Such a dense and cold matter can not be prodused in ordinary heavy ion collisions even by variation of initial energy, A, and impact parameter. The reason for that is the correlation between the heating of the matter and its compression within such an approach.

It is well known that ordinary nuclear matter has a multinucleon component in the form of local($r \sim r_N$[4]) fluctuation (SRC or fluctons)[5,6]. Let's consider the kinematic of the process $A_1+A_2 \rightarrow a+X$, where particle a – meson at central rapidity region with very high $p_t$, close to kinematical limit for the interaction between nucleus $A_1$ and $A_2$. It is clear that for heavy nucleus $A_1$ and $A_2$ and for high energy collisions the probabitity of such a process is negligible. But for He at $E_0 \sim 1$-3GeV/nucleon is not so hopelessly, as it will be shown below. (Later in this article we will talk only about He+He collisions). So, 3-4 nucleons from each colliding nucleus in the form of local multinucleon fluctons interact with each others and produce particle a in the high $p_t$ kinematical region close to kinematical boundary for HeHe collision. He is relatively compact nucleus ($r_{rms} \sim 1.4$fm). Even most conservative estimate for the density of the matter in such a collision leads us to the value of $\rho > 2\rho_0$.

The set of particles forming the system X depends on trigger particle quantum numbers[cum]. The system X will tend to have minimum internal energy when the trigger particle approaches kinematical limit. It means that we will have not only dense but also cold system X in the final state. If the size of the system X is of the order of nucleon size it can be considered as a droplet of dense and cold matter in case of nucleon relative momenta $\delta p < 0.3$GeV and, correspondingly, $T_0 < 50$MeV. Price to be payed for the access into new phase diagram domain is a relatively small size of the droplet and small ( but measurable) cross cection of the process. As for the size of the droplet, one should take into account that the criterion of a medium is l>>r, where l-mean length of free path and r-the size of the system. For the ordinary nuclear matter l~1-2 fm and nucleus heavier than Carbon is usually considered as droplet of nuclear matter. The larger the density the smaller mean length of free path. 6-8 nucleons in the volume of the order of one nucleon volume correspond to the density tens times larger then $\rho_0$, and one can expect l<<1fm. It means one can expect the properties of cold and dense nuclear matter even for droplet size of the order of 1fm.

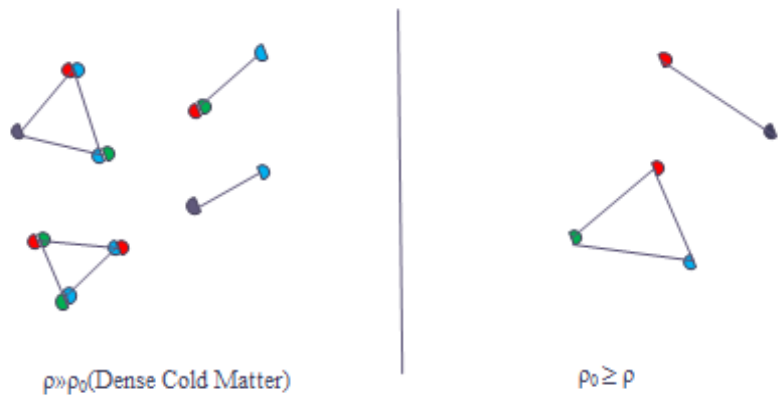

Fig.1. Schematic illustration of nucleon and meson formation process from quarks(color simbols)and antiquarks(black simbols) for ordinary matter(right side) and for dense matter(left).

To understand what properties of the matter we expect to see in the produced droplet let us take into account the deficit of free energy in the produced system and Pauli principle. The solution acceptable for neutron star- the increase of Fermi momentum of nucleons - not exeptable because of the free energy deficit. Extra nucleons must be transformed into other states without significant increase of the effective mass of the droplet. Dibaryons or(and) pentaquarks, if exist, could provide acceptable solution.

This considerations leads us to the proposal to search for pentaquark in the reaction He+He→aX at initial energy ~2-3GeV/nucleon, where a- trigger particle in midrapidity range with maximal possible momentum, and X- a system of N≥6 baryons. Quantum numbers of exotic particles can be changed "by hand" choosing corresponding particle a as the trigger particle. For example? If particle a - negative kaon and Θ+ mass is roughly the summ of nucleon's and kaon's (the difference could be of the order of pion mass or smaller), then such a pentaquark would be the part of system X very likely and might be detected in the decay modes $pK_s$ or $nK^+$. ($m_\Theta - m_N - m_K \sim 0.1 GeV$ is being widely considered for pentaquark candidates).

System X has the same transverse momentum as trigger particle a, and hence, can be kinematically separated from rare spectators. Since the system X is cold, nucleon from Θ+ decay is the fastest one in the system X and combinatoric background is expected to be small. Nuclear matter which could destroy invariant mass peak in heavy ion collisions is practically absent in He+He interactions. To the best of our knowledge such experiments are not realized until now.

To search for dibarions d' with the mass m ~ 2.07GeV[7] the same approach can be used with pion or photon as trigger.

### 3 . Cross sections estimate.

There have been measured by CLAS[8] two and three nucleon SRC probability for He and C; these probabilities are shown in the table. To estimate 4N SRC probability one can take into account two facts. 3N SRC probability is several times smaller than 2N probability squared. Proton-proton SRC probability is at least one order of magnitude suppressed with respect to np SRC[9] (fig.2). Since 4N SRC has at least 2 pairs of identical nucleons let us consider for estimate $a_{4N} \sim (a_{3N})^2/a_{2N}$ (right column of table). Based on probabilities shown in the table one can estimate the fraction of flucton-flucton interactions in $^4He^4He$ collisions as $0.18*10^{-4}$, $0.46*10^{-6}$, $1.2*10^{-8}$ for the total number of nucleons involved into flucton-flucton collisions $N_1+N_2=6,7,8$ accordingly. Let us consider for the first estimate ideal detector with full particle ID in angular-momentum region. If He beam intensity would be $1.5*10^9$ sec$^{-1}$, He target efficiency 0.2 and exposition as long as $3*10^6$ sec, one can accumulates $2*10^{10}$, $5*10^8$ and $10^7$ events for the total number of nucleons involved into flucton-flucton collisions $N_1+N_2=6,7,8$ accordingly. It is a large statistic, but for ideal detector.

Only small fraction of flucton-flucton interaction provides dense cold droplet in the final state. The main fraction is simple multinucleon system with secondary particles distributed over practically all available phase space of the reaction. This fraction depends strongly on the parameters of the droplet and can be estimated only roughly. It is clear that the larger initial energy the smaller the events fraction with dense cold droplet in the final state. On the other hand, the larger initial energy the larger droplet transverse momentum and, consequently, the background conditions. It seems that optimal initial energy range for the proposed measurements $T_0 \sim$ 1-2GeV/nucleon, and it is really possible to get information for droplets with $N_1+N_2 \sim$ 6-7.

### 4 . Trigger and kinematical cooling.

Proposed in section 2 trigger not only selects flucton-flucton interactions, but also selects the final state with relatively small internal energy. The effect can be named as kinematical cooling. The physical reason for the kinematical cooling is the fast (exponential) decreasing of probability of cumulative process (in our case-production of the baryonic droplet) with the increasing of the minimum mass of fluctons ($m_0 \sim \exp\{-T_n/T_*\}$, where $T_n$ -kinetic energy of the nucleon in the droplet and $T_*$ - the slope parameters ~ 60MeV). Absolute minimum in the flucton mass corresponds to the zero internal energy of the baryonic system and, and consequently, zero phase space of the reaction (for nonrelativistic system $dS=d^4p_1...d^4p_n\delta(p_i^2-m_i^2)\delta^4(\sum_1^n p_i - P_n) \sim T_n^{(3n-5)/2}$). Maximum for $dS *m_0^2$ would be for nucleon kinetic energy of the order of $T \sim (3n-5)T*/2n \rightarrow 3T*/2 \sim 100$ MeV, which corresponds to nucleons relative momenta within the droplet $p \sim 0.45$ GeV/c.

The efficiency of proposed trigger can be estimated from data accumulated by FLINT collaboration[10]. FLINT obtained ~ $10^3$ events at $Q_1+Q_2>4$ with photon trigger. From simulations we know that effective number of nucleons in the droplet exceeds the minimum one for the value about 0.5-1.0. In case of photon trigger there is additional uncertainty because of unknown mechanism of photon production. Photon can be direct one or from pion decay. In the last case the second photon carry on a small, but not negligible fraction of pion energy. For that reason effective value $N_1+N_2$ for this experimental data can be estimated as large as ~5-6. It was done during ~50 hours exposition with detector acceptance ~$10^{-2} *4\pi$ and CBe interaction rate ~$2*10^6$ sec$^{-1}$. It looks quite real to obtain data for 1000 hours exposition with interaction rate ~$2*10^8$ sec$^{-1}$ with ten times larger detector acceptance and different types of triggers ($\gamma,\pi^{\pm},K^{\pm}$). Even for HeHe interactions (~0.1 supresion with respect to CBe) it provides estimated number of accumulated events ~$2*10^8$ for $N_1+N_2$ ~6 at initial energy 2 GeV/nucleon. Total numbers of flucton-flucton interactions at the same conditions two order of magnitude larger($2*10^{10}$-section 3).

To summarize this section one can conclude:

1) When applicable, such trigger suppresses statistics by two order of magnitude and then makes off-line analysis much more easy.

2) Kinematical cooling provides relative momenta scale within droplet (~0.45GeV/c) comparable and not smaller than estimated relative momentum in the dense cold matter droplet (~0.3GeV/c). It means that trigger does not significantly suppresses events under study.

|  | $a_{2N}$, % | $a_{3N}$, % | $(a_{2N})^2$, % | $(a_{3N})^2/a_{2N}$, % |
|---|---|---|---|---|
| $^3$He | 8.0±1.6 | 0.18±0.06 | 0.64 | 0.004 |
| $^4$He | 15.4±3.3 | 0.42±0.14 | 2.4 | 0.011 |
| $^{12}$C | 19.3±4.1 | 0.55±0.17 | 3.7 | 0.016 |

Table 1. 2N and 3N SRC probabilities according to [Egiyan] and 4N SRC probability estimate as $(a_{3N})^2/a_{2N}$

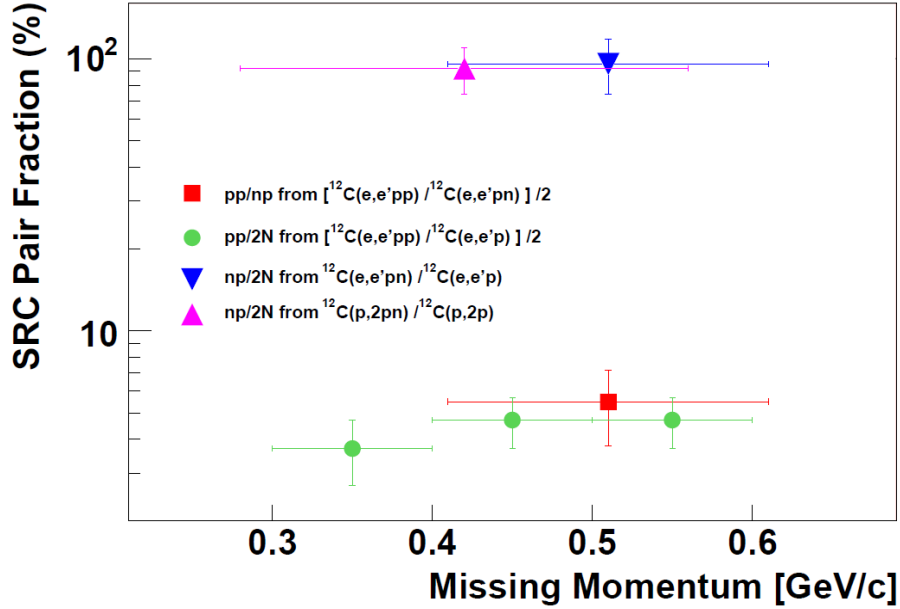

Fig.2. The fractions of correlated pair combinations in carbon as obtained from the (e,e'pp) and (e,e'pn) reactions, as well as from previous (p,2pn) data[isotopic].

## 5 . From femtoscopy to femtotechnologies.

Experimental confirmation of the formation of cold and dense baryonic matter droplet in the events selected by trigger should be preceded with the search for exotic. "Cold baryonic droplet" experimentally means a bump in the nucleon-trigger correlation function with relatively small nucleon-nucleon mean relative momentum within this bump. The density of the droplet depends on space – time interval between its constituents. The latter is usually measured by Kopylov-Podgoretsky method, now frequently called as femtoscopy due to characteristic scale of measured size ~1-10fm. Starting from femtoscopy, which is only passive test for space-time parameters of the object under investigation, we achieve in our study the active control of the of the process on the space scale of the order of fm.

The choice of trigger particle specie determine minimal configuration of baryonic droplet. One need K$^-$ trigger to search for Θ+ because of such a trigger provides minimum droplet configuration with one strange antiquark, while pion or photon trigger would be more preferable to search for di-baryon d'. For the same reasons K$^+$ trigger would be a good choice to search for multistrange hyperonic system.  More common statement that the choice of trigger particle specie controls quantum numbers of produced baryonic droplet.

On the other side, the trigger particle momentum variation change the minimum number of nucleons within droplet, and, consequently, the initial density of the droplet. The density variation provides transition from one phase diagram domain to another and, consequently, provides access to new state of matter through the phase transitions.

All these processes proceed at the space-time scale of the order of 1-10 fm, and one can say about an embryo of future femtotechnologies in such experiments.

### 6. Conclusions.

We propose the way to create and to investigate a dense cold matter droplets in the laboratory. The reality of this approach are argued. Estimated possible statistic is large enough for detail study of the properties of such a matter.
Mechanism of kinematical cooling of the droplet is clarified.
It is shown that in such a way new upper limit of cross sections is accessible for the study of different kind of exotic.
Different types of trigger(selection criteria) are proposed to search for different kind of exotic.

We are grateful to S.S.Shimanskiy for useful discussions and remarks.

### Referencies.